\documentclass[aps,prl,twocolumn,preprintnumbers,showpacs,amsmath,amssymb]{revtex4-1}
\usepackage[colorlinks=true,linkcolor=blue,citecolor=blue,urlcolor=blue]{hyperref}
\usepackage[utf8]{inputenc}
\usepackage{graphicx}
\usepackage{verbatim}
\usepackage{bm,footnote}

\newcommand{\gp}{\ensuremath{g'}} 
\newcommand{\lp}{\ensuremath{l'}}
\newcommand{\fp}{\ensuremath{f'}}
\newcommand{\eps}{\ensuremath{k}}
\newcommand{\psic}{\ensuremath{\psi^\mathrm{c}}}
\newcommand{\abs}[1]{\ensuremath{\left| #1 \right|}}
\renewcommand{\vec}[1]{\ensuremath{{\textbf{\em #1}}}}
\newcommand{\mat}[1]{\ensuremath{{\rm \bf #1}}}
 
\newcommand{\expe}[1]{\ensuremath{{\rm e}^{#1}}}

\begin{document}

\title{Amplitude death in networks of delay-coupled delay oscillators}

\author{Johannes M. H\"ofener$^1$, Gautam C. Sethia$^2$ and Thilo Gross$^3$}
\address{%
$^1$Biological Physics Section, Max-Planck-Institut f\"{u}r Physik komplexer Systeme, N\"{o}thnitzer Stra\ss e 38, Dresden 01187, Germany\\
$^2$Institute for Plasma Research, Bhat, Gandhinagar 382 428, India\\
$^3$University of Bristol, The Merchant Venturers School of Engineering, Department of Engineering Mathematics, Bristol, UK}%

\date{\today}

\begin{abstract}
Amplitude death is a dynamical phenomenon in which a network of oscillators settles to a stable state as a result of coupling. 
Here, we study amplitude death in a generalized model of delay-coupled delay oscillators. We derive analytical results for degree homogeneous networks that show that amplitude death is governed by certain eigenvalues of the network's adjacency matrix. 
In particular these results demonstrate that in delay-coupled delay oscillators amplitude death can occur for arbitrarily large
coupling strength $\eps$. In this limit we find a region of amplitude death, which occurs already at small coupling delays that scale with 1/$\eps$. We show numerically that these results remain valid in random networks with heterogeneous degree distribution. 
\end{abstract}

\maketitle


\section{Introduction}
Coupling individual oscillators can lead to amplitude death, the cessation of oscillations due to stabilization of a stationary state\cite{Atay2010}. 
This phenomenon has been observed in chemical oscillators \cite{Bar-Eli1985}, electronic circuits \cite{Reddy2000}, thermo-optical oscillators \cite{Herrero2000}, and coupled lasers \cite{Prasad2003}. 
Amplitude death can be beneficial and is exploited for instance in feedback control applications \cite{Konishi2008}, 
but can also be detrimental and even lethal if it occurred in interacting cardiac cells \cite{Atay2006JDE}.

In mathematical models, amplitude death is commonly studied by investigating the conditions under which an unstable steady state of an isolated system is stabilized by coupling. It was found that for amplitude death to occur, either the natural frequencies of the coupled systems need to be sufficiently disparate \cite{Yamaguchi1984,Shiino1989,Aronson1990,*Ermentrout1990,Mirollo1990} or the coupling needs to be time-delayed \cite{Reddy1998}.

While amplitude death can be observed already in systems of two oscillators, richer behavior is observed when complex networks of oscillators are considered. In such networks the nodes represent individual oscillators, whereas the links represent coupling terms. A central question is then how the dynamics are affected by network topology, the specific structure of nodes and links. For a given network of $N$ oscillators this topology can be captured by the adjacency matrix $\mat{A}$, a $N \times N$ matrix with $A_{ij}=1$ if a link exists from node $j$ to node $i$ and $A_{ij}=0$ otherwise. A related matrix known to affect many network properties is the network Laplacian $\mat{L} = \mat{A} - {\rm \bf D}$. Here, $\mat{D}$ is a diagonal matrix, with $D_{ii} = k_i$, the degree of node $i$, i.e.~the number of nodes to which the focal node is connected.   

Delay induced amplitude death is often studied in systems of coupled limit-cycle oscillators, such as Stuart-Landau oscillators. Studies on globally connected networks \cite{Reddy1998,*Reddy1999} and rings \cite{Dodla2004} showed that amplitude death occurs inside islands in the parameter space of coupling strength and delay. While distributed delays seem to enlarge these islands \cite{Atay2003PRL}, gradient instead of diffusive coupling has been found to suppress amplitude death in rings \cite{Zou2010}. Further, transients between regimes of partial and complete amplitude death have been studied by numerical simulations \cite{Yang2007}. Also the effect of delayed self-feedback has been studied in a single and two delay-coupled Stuart-Landau oscillators \cite{Reddy2000PhysD, DHuys2010}. 

Delay-induced amplitude death has been explored in detail in discrete-time maps \cite{Atay2004,*Atay2006SIAM,Masoller2005,Gong2008,Ponce2009} and in systems of oscillators modeled by multi-dimensional ordinary differential equations, such as the R\"{o}ssler and Lorenz oscillators \cite{Atay2003PhysD,Atay2006JDE,Konishi2004,*Konishi2005,Prasad2005}. For a model of coupled maps, it was shown that the dynamics is governed by the largest eigenvalue of the network's Laplacian \cite{Atay2006SIAM}. A similar result was shown in time-continuous system \cite{Atay2006JDE}.

Previous studies have shown impressively that analyzing amplitude death in simple systems, such as simple oscillators or maps can yield a deep understanding, whereas coupling more complex systems such as R\"{o}ssler or Lorenz oscillators points to additional complexities. 
A middle way is perhaps offered by using delay oscillators, such as Mackey-Glass \cite{Mackey1977,Farmer1982} and the Ikeda \cite{Ikeda1987} oscillators. Because even a single-variable delay-differential equation (DDE) constitutes an infinite dimensional dynamical system, a single DDEs can show sustained oscillations, quasi-periodicity, and chaos. Studying amplitude death in delay-coupled systems of such delay oscillators offers the opportunity to consider a network of relatively complex coupled system, while keeping the equations concise.  

Although a whole zoo of different synchronization types have been found for systems of coupled delay-oscillators without coupling delay \cite{Pyragas1998,Voss2000,Zhan2003,Li2004,Shahverdiev2005,Chen2007}, delay-coupled delay oscillators, have been less explored. One exception is \cite{Konishi2008} where a system of two mutually delay-coupled delay oscillators was investigated numerically, but more 
complex coupling topologies were not considered. 

A model for complex networks of general scalar delay-coupled delay oscillators was proposed in two previous 
publications of the present authors \cite{Hoefener2011,*Hoefener2012}. The latter paper highlighted that certain meso-scale structures in networks have a distinct impact on the network-level dynamics, whereas the former discussed the effect of the degree distribution, the probability distribution of node degrees.
For degree-homogeneous networks (DHONs), where every node is connected to the same number of of other nodes, the dynamics of delay-coupled delay oscillators were studied analytically. This revealed the bifurcation lines at which the dynamics of the system change qualitatively, and linked them to certain eigenvalues of the networks adjacency matrix.  
For degree-heterogeneous networks (DHENs), numerical explorations showed qualitatively similar dynamics, with minor corrections.  

In the present paper, we use the previously proposed model to study amplitude death. For this purpose, we focus on a regime where the uncoupled systems show non-stationary dynamics. Then, we study the parameter space of coupling strength and coupling delays to find stable areas, which correspond to regions of amplitude death. For DHONs, we find that there exists a region of amplitude death for coupling delays larger than a certain threshold value, which approaches zero as $1/(dk)$ where $d$ is the common degree of nodes and the coupling strength $k$ approaches infinity. By using a numerical sampling method, we confirm this result for degree-heterogeneous networks, where $d$ is now the mean degree of the network. This shows that in the limit of large coupling strength the dynamics is governed by a global property of the network, while all other topological properties become negligible.

\section{Model}
We consider networks of $N$ nodes $i$, carrying a dynamical state $X_i$, which can represent for instance an ecological population or the abundance of RNA molecules in a genetic oscillator. The variable is subject to gain and loss terms, where the loss is 
instantaneous, while the gain term of $X_i$ at time $t$ depends on the value of $X_i$ at time $t-\tau$.  
Furthermore, $X_i$ continuously diffuses to topological neighbors causing an additional loss. 
The exported material arrives at the respective neighboring node $j$ after a travel time delay $\delta$. 
Thus, the dynamics is governed by
\begin{equation}\label{eq:DDE}
\dot{X}_i=G(X_i^{\tau})-L(X_i)+\sum_j \left(A_{i j} F(X_j^{\delta})-A_{j i}F(X_i)\right),
\end{equation}
where $G$, $L$ and $F$ are positive functions describing growth, loss, and coupling, respectively.
Instead of restricting these functions to specific functional forms, we consider a general model comprising the whole class of models that includes well-studied examples such as the delay-coupled Mackey-Glass and the Ikeda systems.

Here we only consider networks for which the number of outgoing links $d_i^\mathrm{out}$ equals the number of incoming links $d_i^\mathrm{in}$ for each node $i$, such that $d_i^\mathrm{out}=d_i^\mathrm{out}=:d_i$, where $d_i$ is the degree of node $i$. 
We note that this class includes among all bidirectionally coupled networks. 
For all networks in the class considered, a homogeneous steady state $X_i*=X^*$ exists, for each steady state $X^*$ of the isolated system, which satisfies $G(X^*)=L(X^*)$.

We apply the method of generalized modeling \cite{GMPRE,GMSCIENCE}, which analyzes the dynamics of general system by a direct parametrization of the Jacobian matrix ${\rm \bf J}$, with $J_{ij} = \partial \dot{X}_i / \partial X_j$ in $X^*$. This matrix constitutes a local linearisation and thus governs the dynamics of the system close to a steady state under consideration \cite{Kuznetsov}. In order to express the Jacobian in terms of interpretable parameters, we normalize the system to an arbitrary homogeneous steady state ${X_i}^*$ and introduce normalized variables $x_i=X_i/X^*$ and normalized functions $f(x_i)=F(x_i X^*)/F(X^*)$, denoted by lower case symbols. 
Using $X^*=X^{\tau *}=X^{\delta *}$, we rewrite Eq.~\eqref{eq:DDE} as
\begin{equation}\label{eq:DDE_norm}
\dot{x}_i=\alpha(g(x_i^{\tau})-l(x_i))-d_i \beta f(x_i)+\beta\sum_j A_{i j} f(x_j^{\delta}).
\end{equation}
The quantities $\alpha=G(X^*)/X^*=L(X^*)/X^*$ and $\beta=F(X^*)/X^*$ are unknown constants and can thus be interpreted as  parameters of the Jacobian. These parameters have the unit of inverse time and describe turnover rates. In the following we set $\alpha=1$ by a time scale normalization.

In order to analyze the stability of the homogeneous steady state, we proceed as in \cite{MacDonald1989} and assume that, close to the steady state, the DDE has exponential solutions $\vec{y}(t)=\vec{y}(0)\exp(\lambda t)$, where $\vec{y}(t)=\vec{x}(t)-1$ is a small perturbation from the steady state at $\vec{x}=\vec{1}$. Using this ansatz together with the linearization of Eq.~\eqref{eq:DDE_norm} the Jacobian matrix has the entries  
\begin{equation}
\label{eq:Jac}
\begin{array}{r c c l}
J_i^\mathrm{d}&:=&\gp\exp(-\lambda\tau)-\lp-d_i\beta\fp    & \text{for } i=j\\ 
J^\mathrm{o}&:=&\beta\fp\exp(-\lambda\delta)               & \text{for } A_{i j}=1\\
0   & &                                                    & \text{otherwise}
\end{array}
\end{equation}
where $\lambda$ is an eigenvalue of the Jacobian and thus a solution of the characteristic polynomial $P(\lambda)=\det(\lambda\mat{I}-\mat{J})$. The quantities $\gp,\lp,\fp$ denote derivatives of the normalized functions in the steady state, or equivalently logarithmic derivatives of the original functions, i.e.~$\gp=\left.\partial g / \partial x \right|_{1}=\left.\partial \log{(G)} / \partial  \log (X)\right|_{*}$. Although we do not specify the functions or the steady state under consideration, these derivatives are formally constants and can thus be interpreted as unknown parameters of the system. Closer inspection reveals that these so-called elasticities are generally easily interpretable in the context of applications and have a number of additional benefits\cite{GMPRE}. Since $\fp$ only appears in products together with $\beta$, we introduce the effective coupling strength $k:=\beta\fp$.

The steady state under consideration is asymptotically stable if all roots of the characteristic polynomial have negative real-parts and it is unstable if at least one root has a positive real part \cite{Atay2010}. In DDEs the explicit appearance of $\lambda$ in the Jacobian turns the characteristic polynomial into a transcendental function, which may have an infinite number of solutions. To make progress analytically it is therefore advantageous to directly compute the bifurcation points in parameters space at which eigenvalues acquire positive real parts. This is general simpler than computing the spectrum for a single parameter set. In numerical explorations the computation of the spectrum can be avoided by checking for positive eigenvalues with a method based on Cauchy's argument principle \cite{Luzyanina1996,Hoefener2011}. 

\section{Bifurcations of degree homogeneous networks}
In a previous publication \cite{Hoefener2011}, we demonstrated that the characteristic equation $P(\lambda)=0$ of DHONs can be decomposed into $N$ independent equations, each of which corresponds to a single eigenvalue of the adjacency matrix. 
In this section we briefly recapitulate this derivation that will form the basis for our discussion of amplitude death below. 
Because the treatment involves both the eigenvalues $\lambda_i$ of the system's Jacobian matrix, and the eigenvalues $c_i$ of the system's adjacency matrix, we refer to $\lambda_i$ as dynamic eigenvalues and to $c_i$ as topologial eigenvalues.  

In a DHON all diagonal elements $J_i^\mathrm{d}$ are identical. This allows us to substitute $\lambda-J^\mathrm{d}$ in the characteristic polynomial $P(\lambda)$ by $z$, which yields $P'(z)=\det(z\mat{I}-J^\mathrm{o}\mat{A})$. The roots of $P'(z)$ are given by $c_i J^\mathrm{o}$, where $c_i$ denotes one of the $N$ topological eigenvalues. The back-substitution yields $N$ independent equations,
\begin{equation}\label{eq:EV_DHON}
\lambda=J^\mathrm{d}(\lambda)+c_i J^\mathrm{o}(\lambda).
\end{equation}
We note that the topological eigenvalues $c$ are in general complex, so that $c=|c|\expe{i\psic}$, where $\psic$ is the complex phase of the topological eigenvalue. 

For determining the stability of the system, we recall that the stability changes at bifurcation points where a dynamical eigenvalue crosses the imaginary axis and becomes purely imaginary, so that $\lambda=i \omega$. By separating the real and imaginary part of Eq.~\eqref{eq:EV_DHON} we obtain
\begin{eqnarray}
\label{eq:Re}0&=&\gp\cos(\phi)-\lp-d\eps+|c|\eps\cos(\psi),\\
\label{eq:Im}\omega&=&-\gp\sin(\phi)-|c|\eps\sin(\psi),
\end{eqnarray}
with $\phi=\omega\tau$ and $\psi:=\omega\delta-\psic$.

The equation system contains three unknown variables, $\phi$, $\psi$ and $\omega$. Given a solution triplet $(\phi,\psi,\omega)$, we can find another solution $(-\phi,-\psi,-\omega)$. The solutions with negative $\omega$ trace the same bifurcation lines as those with positive $\omega$. We can thus restrict the analysis to solutions $\omega\geq 0$ without missing bifurcations. Further, we are only interested in solutions with $\tau>0$, and therefore $\phi>0$. By choosing $\phi$ as a free parameter, we can find a parametric representations of the bifurcation lines, but need to distinguish three cases depending on the value of the topological eigenvalue $c$. 

For the case $c=0$ the Eqs.~(\ref{eq:Re},\ref{eq:Im}) are independent of $\psi$ and $\delta$, respectively. Thus, the bifurcations are vertical lines in the $(\eps,\delta)$-plane, with 
\begin{equation}\label{eq:beta_c0}
\eps=\frac{1}{\tau}\frac{h}{d},
\end{equation}
where $h:=(\gp\cos(\phi)-\lp)\tau$ and $\phi$ has to satisfy $f(\phi):=\phi+\gp\tau\sin(\phi)=0$. The equation $f(\phi)=0$ may generally have several solutions. However, for the parameters used throughout this paper the solution is unique.

For the case $|c|=d$, we find
\begin{equation}\label{eq:beta_cd}
\eps=\frac{1}{\tau}\frac{f^2+h^2}{2dh}.
\end{equation}
In order to obtain positive solutions, $h(\phi)$ needs to be positive. Therefore, we have to restrict $\phi$ to the intervals $I^\eps_r=[\phi^\eps_r,\overline{\phi^\eps_r}]$, with $\phi^\eps_r=2\pi r + \phi^\eps$, $\overline{\phi^\eps_r}=2\pi(r+1)-\phi^\eps$ and $\phi^\eps=\cos^{-1}(\lp/\gp)$. The parameter $r$ that appears here is the first of two integer parameters that we introduce to enumerate the bifurcation lines.  
 
Finally, for the case $0<|c|<d$, we find
\begin{equation}\label{eq:beta}
\eps_\mathrm{1,2}=\frac{1}{\tau}\frac{dh\pm\sqrt{a}}{d^2-\abs{c}^2},
\end{equation}
with
\begin{equation}
\label{eq:a}a=d^2h^2-(d^2-|c|^2)(f^2+h^2).
\end{equation}
For obtaining real and positive solutions, we require not only $h(\phi)>0$ but also $a(\phi)>0$. This imposes additional constraints on $\phi$ that can be calculated numerically. In contrast to the case $|c|=d$, valid solutions can only be found inside a finite number of intervals $I^\mathrm{\eps}_r$.

We remark that the case $|c|>d$ does not need to be considered because the eigenvalues $c$ of the adjacency matrix cannot exceed the highest node degree, which is $d$ in DHONs.  

For the cases above, we compute the corresponding values of $\delta$ at which the bifurcation occurs from Eqs.~(\ref{eq:Re},\ref{eq:Im}). We find
\begin{align}
\label{eq:delta}\delta^\mathrm{L,R}&=\frac{\psi^\mathrm{L,R}+\psi_{\rm c}+2\pi s}{\phi}\tau,\\
\label{eq:psi}\psi^\mathrm{L,R}&=\pm\cos^{-1}\left(\frac{d}{|c|}-\frac{h}{|c|\eps}\right),
\end{align}
where $s$, is the second integer parameter introduced to enumerate solution branches. 
Furthermore, the indices L, respectively R, are introduced to denote the positive, respectively negative, sign branch of $\psi$. 
To obtain physical solutions we consider the L-branch for all $f(\phi)<0$ and the R-branch otherwise.
Evaluating these, we find positive values of $\delta$ for non-negative integers $s$.

One implication of Eq.~\eqref{eq:psi} is revealed when we consider that the arcus cosine needs to be smaller or equal to one. This is only possible if $\eps\le\eps^\mathrm{max}$ with
\begin{equation}
\eps^\mathrm{max}=-\frac{l+g}{d-|c|}.
\end{equation}
Thus, bifurcation lines corresponding to topological eigenvalues $c$ with $|c|<d$ can only be found for finite values of $\eps$. Hence, only eigenvalues with $|c|=d$ can affect the stability for sufficiently large coupling strength, independently of the coupling topology.


\section{Amplitude death in degree homogeneous networks}
Having derived the results in the previous section, we now turn to the analysis of amplitude death. In the parameter range $\gp<-\lp<0$, we find that an isolated node is unstable if $\tau>\tau^*$, where $\tau^*=\tau^*_{r=0}$ and 
\begin{equation}
\tau^*_r=\frac{\phi^\eps+2\pi r}{\sqrt{\gp^2-\lp^2}}.
\end{equation}
The isolated system thus exhibits non-stationary (e.g.~oscillatory) dynamics when the reproductive delay $\tau$ is chosen sufficiently large.  
The stationary solutions can then potentially be stabilized by coupling the oscillators. In general stabilization will depend on the coupling topology, coupling delays $\delta$, and the coupling strength $k$. 
In the following we seek to identify the effect of the coupling topology on the areas in the $(\eps,\delta)$-space where the homogeneous steady state is stable, such that amplitude death can occur.

\begin{figure}
\centering
\includegraphics{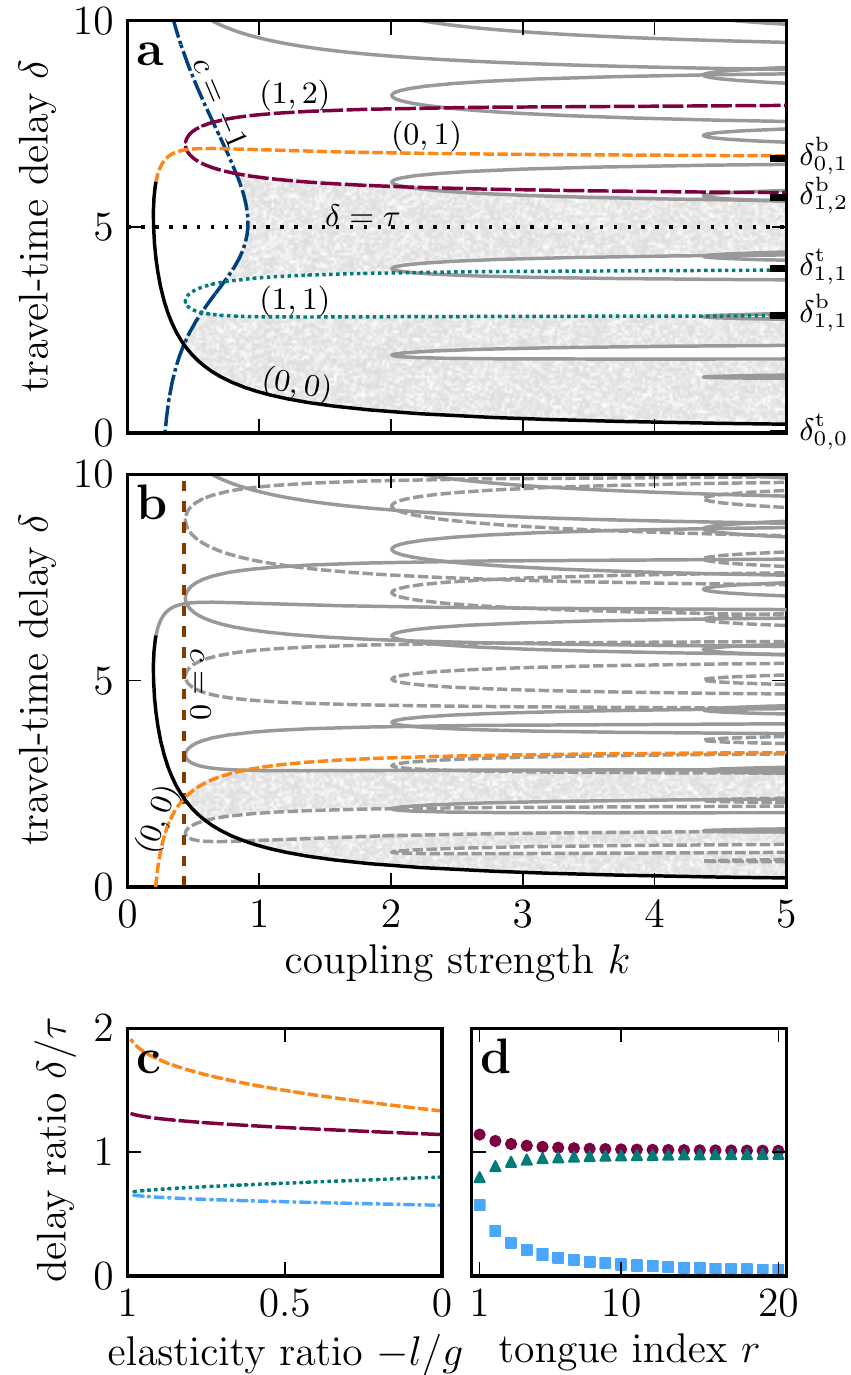}
\caption{Bifurcation diagrams indicating the dynamics for a fully connected network of 3 nodes (a) and a ring of four nodes (b). 
Lines mark bifurcation points corresponding to different topological eigenvalues $c_i$ (see text). Labels of the form $(r,s)$ on some of the 
lines state the corresponing indices $r$ and $s$ enumerating the analytical solution branches. 
For the limit of high coupling strength the bifurcation lines asymptotically approach limiting values, some of which are indicated on the right border of a) by symbols of the form $\delta_{r,s}^\mathrm{b,t}$, where $b$ and $t$ denote bottom and top branches, respectively.
Gray dots mark stable states that were found numerically by uniformly sampling the ($k$,$\delta$) plane.   
The two lower plots (c,d) show the dependence of limiting values $\delta_{r,s}^\mathrm{b,t}$ on the ratio of local parameters $-l/g$ (c) and the dependence of $\delta_{r,r+1}^\mathrm{b}$, $\delta_{r,r}^\mathrm{t}$ on the index $r$. 
Parameters: $\gp=-1,\lp=0,\tau=5$.}   
\label{fig:BifLines}
\end{figure}

Our analytical treatment above has identified a connection between the topological eigenvalues and the dynamical stability. 
For illustration of the analytical results we consider two specific topologies: A fully connected networks of 3 nodes and a ring of four nodes. 

Let us start with the fully connected network (Fig.~\ref{fig:BifLines}a). 
All such networks have a topological eigenvalue $c=d=N-1$ and a $d$-fold degenerate topological eigenvalue $c=-1$.
For $c=-1$ computation of the bifurcation lines only yields physical solutions for the $r=0$ branch.
In this branch different values of $s$ generate different segments of bifurcation lines, which connect such that a single long bifurcation line running from $\delta=0$ to $\delta=\infty$ is formed.
Because $|c|<d$, all bifurcation points in this line occur at finite values of the effective coupling strength $\eps$.   

For the remaining eigenvalue $c=d$, the situation is more complex. Here, computation of the bifurcation points yields a family of separate tongues corresponding to different values of $r$ and $s$.  Similar families of bifurcation lines have been observed in other delay systems \cite{SchollTwoDelays}. In the present paper, each of the tongues consists of two segments, to which we refer as the bottom and top part respectively. An exceptional solution is again the case $r=0$. Here, the lower boundary of the tongue is formed by the \emph{top} branch of the $(r=0,s=0)$ solution, while the upper boundary is formed by the \emph{bottom} branch of the $(0,1)$ solution. One can thus think of the $r=0$ case as an inside-out tongue, which provides the right intuition for understanding the results below. 

Let us now turn to the ring network (Fig.~\ref{fig:BifLines}(b)). All even rings are bipartite networks, which means that it is possible to color the nodes in two colors such that no node has a topological neighbor of the same color. Bipartite degree homogeneous networks have topological eigenvalues at $c=d$ and $c=-d$, additionally the 4-ring has a two-fold degenerate eigenvalue at $c=0$, which results from symmetry.  
For the topological eigenvalue $c=0$ only the ($r=0$,$s=0$) solution corresponds to a physical bifurcation line. This line occurs at a constant value of the coupling strength. The eigenvalues with $c=d$ and $c=-d$ each generate a family of tongues that are similar to those in the fully connected network. However, the tongues with $c=-d$ are shifted such that they are centered on the gaps between the tongues with $c=d$. From the corresponding eigenvectors one can see that the family of tongues with $c=d$ corresponds to an instability with respect to in-phase oscillations, whereas the tongues with $c=-d$ correspond to anti-phase oscillations. It is intuitive that this instability arises from the eigenvalue $c=-d$ which is directly linked to bipartiteness, because none-bipartite networks could not sustain anti-phase oscillations.   

The analytical results above can be confirmed by a numerical sampling procedure, in which we pick parameter sets uniformly from the $(k,\delta)$-plane and evaluate their stability numerically \cite{Hoefener2011}. For visualization we plot only those points that are found to be stable. The results shown in Fig.~\ref{fig:BifLines} shows that such stable parameter combinations are found only in certain regions that are sharply delineated by the theoretically predicted bifurcation lines. The figure shows that stability requires that the respective point lies inside all tongues with $r=0$ but outside all tongues with $r \neq 0$, which confirms the inside-out nature of the tongues with $r=0$ mentioned above. Thus for stable points, the bifurcation lines corresponding to $r=0$ impose a minimum coupling strength, whereas the solutions with $r\neq 0$ impose a maximum coupling strength.  

We note that the $c=0$ topological eigenvalue never causes a bifurcation of a stable solution.
This eigenvalue gives rise to a physical branch with $r=0$ and thus potentially imposes a lower limit for the coupling strength in stable states. However, this limit is always below a higher limit imposed by the $r=0$ branches of other eigenvalues, and thus has no direct relevance for stability. This conforms to the general observation that only the largest positive and the smallest negative topological eigenvalue cause bifurcations that border stable regions. For DHONs it is probably possible to prove this rigorously, but the proof is beyond the scope of the current paper. Nevertheless, it is interesting to note that in all DHONs the largest positive eigenvalue is $c=d$, whereas the smallest negative eigenvalue is maximal for fully connected networks ($c=0$) and minimal for bipartite networks ($c=-d$). This illustrates why amplitude death is most likely in the fully connected network and least likely in bipartite networks. All other networks fall between these two extreme cases, which motivated our choice of examples.    

 
\section{Strong coupling limit}
In the following, we focus on large coupling strength $\eps$. Above we showed already that only those bifurcation lines that correspond to topological eigenvalues with $|c|=d$ can extend to arbitrarily large coupling strength. Except in bipartite networks, this condition is only satisfied for the eigenvalue $c=d$. In the following we thus focus solely on this eigenvalue.

In the limit $\eps \to \infty$, the bifurcation lines approach constant values of the coupling delay $\delta_{r,s}^\mathrm{b,t}$, respectively, where t and b denote values for top and bottom branches. For calculating $\delta_{r,s}^\mathrm{b,t}$, we study Eq.~\eqref{eq:beta_cd} and find that $\eps$ becomes infinitely large if $\phi$ approaches $\phi^\mathrm{\eps}_r$ and $\overline{\phi^\mathrm{\eps}_r}$. Using Eq.~\eqref{eq:delta} we obtain 
\begin{equation}\label{eq:delta_tb}
\delta^\mathrm{t}_{r,s}=\frac{2\pi s +\psi^\mathrm{c}}{2\pi r +\phi^\eps}\tau,\qquad
\delta^\mathrm{b}_{r,s}=\frac{2\pi s +\psi^\mathrm{c}}{2\pi (r+1) -\phi^\eps}\tau.
\end{equation}
We note that both equations scale linearly with $\tau$, such that all tongues scale with the internal delay of the oscillators. Furthermore, recalling that $\phi^\eps=\cos^{-1}(\lp/\gp)$, we see that for $\gp=-\lp$, the bottom and top limits are identical for each bifurcation line $(r,s)$, such that the unstable areas disappear (see also Fig.~\ref{fig:BifLines}(c) for selected lines). 

Let us now investigate the question whether in the limit of large coupling strength amplitude death is still possible. 
Already visual inspection of Fig.~\ref{fig:BifLines}(a) reveals that there are large channels of stability between the tongues, which seem 
to extend to high values of $\eps$. In particular, notable is the \emph{1:1-resonant channel} around $\delta=\tau$ and the bottom channel at small values of $\delta$. As we increase the coupling strength, all stable channels are successively narrowed down as new tongues, corresponding to branches with higher $r$, become relevant. For instance, the 1:1-resonant channel is bordered from below by the top branch of $(r,r)$ and from above by the bottom branch of $(r,r+1)$; the bottom channel is bordered from below by the top branch of $(0,0)$ and from above by the bottom branch of $(r,1)$.


\begin{figure}[ht!]
\centering
\includegraphics{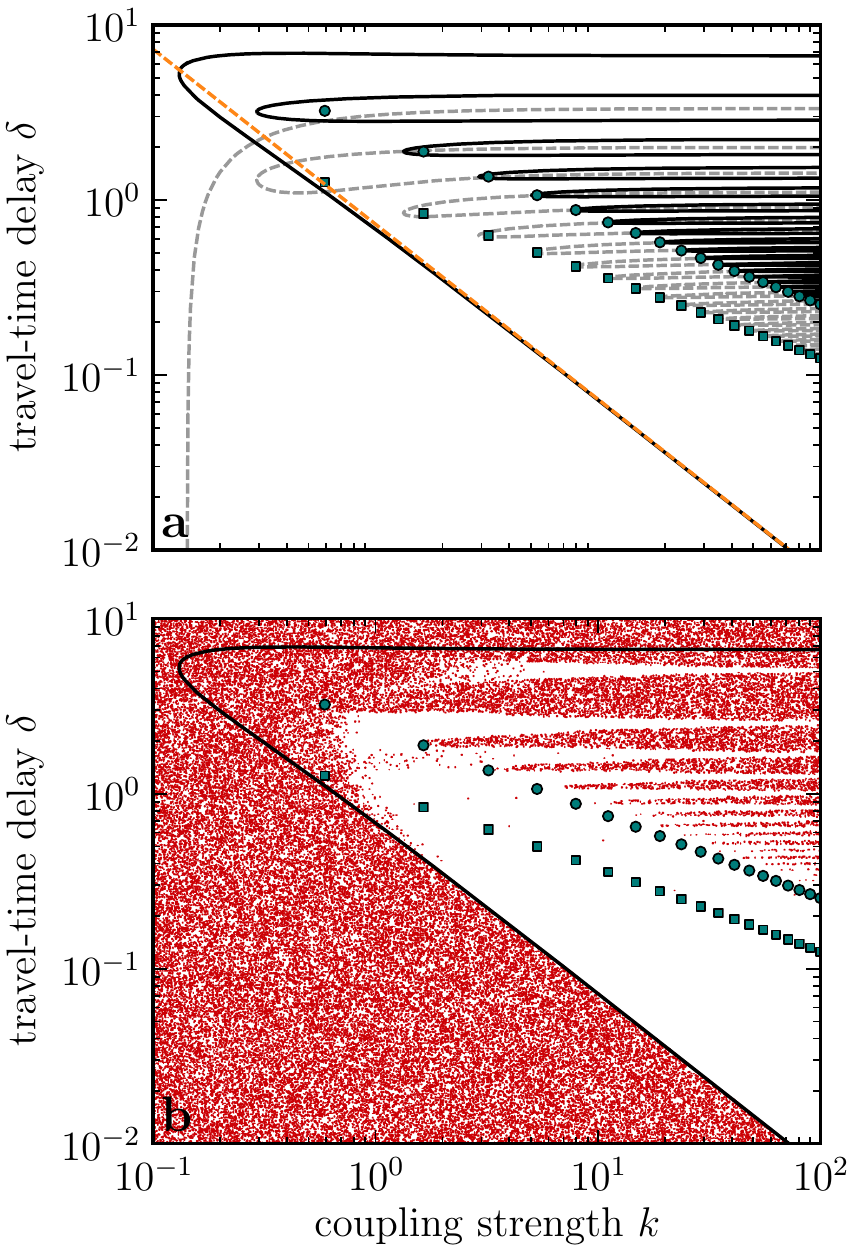}
\caption{Amplitude death for large coupling strength. (a) Bifurcation lines from the eigenvalue $c=d$ (solid lines) and $c=-d$ (dashed lines). The approximate tip positions of tongues have been marked by circles ($c=d$) and squares ($c=-d$).  
The orange dashed line marks the analytical approximation to the bottom boundary of the stable region. For sufficiently large $k$, all degree homogeneous networks are stable between this bifurcation line and the tongues corresponding to $c=-d$. Non-bipartite networks are even stable up the tongues corresponding to $c=d$. (b) Numerical results from uniform sampling of the space shown for random networks with $N=10$ nodes and $K=15$ edges. Red dots mark unstable parameters points, such that the large white regions mark stable areas that correspond to amplitude death. In particular, for sufficiently $k$, no unstable networks are found for $\delta$ between the bottom bifurcation line with ($d=2K/N$) and the tip positions for $c=-d$. Most networks are stable up to the tip positions for $c=d$. In the double logarithmic plot the region of amplitude death growth with increasing $k$ as its boundaries approach power-laws with exponent $-1$ and $-1/2$ respectively, in agreement with analytical predictions. Parameters are $\gp=-1,\lp=0,\tau=5$.}\label{fig:Scaling}
\end{figure}

We now focus specifically on the bottom channel, which comes arbitrarily close to the case of zero coupling delay and thus the well studied case of networks of delay oscillators without coupling delay. We start by considering the lower boundary of this channel by studying the  asymptotic behavior of the $(0,0)$ bifurcation line. We note that in the relevant limit this is given by the L-branch of Eqs.~(\ref{eq:delta},\ref{eq:psi}) with $\phi$ approaching $\phi^\mathrm{\eps}$. In this case $h\to 0$, so that we can approximate Eq.~\eqref{eq:psi} by using $\cos^{-1}(1-x)\approx\sqrt{2|x|}$. With $\tau^*=\phi^\eps/\sqrt{\gp^2-\lp^2}$ and $\phi^\eps=\cos^{-1}(\lp/\gp)$, we find the critical coupling delay
\begin{equation}\label{eq:delta_approx}
\delta=\left(\frac{\tau}{\tau^*}-1\right)\frac{1}{d\eps},
\end{equation}
where we dropped the indices for simplicity. 
This result shows that the critical $\delta$, which marks the lower boundary of the bottom channel, is inversely proportional to the effective coupling strength $d\eps$.

For finding the critical $\delta$ that marks the upper boundary of the bottom channel, we approximate the tip positions of the tongues $(r,1)$ in the limit of large $r$. In this case we can approximate Eq.~\eqref{eq:beta_cd} with
\begin{equation}\label{eq:beta_approx}
k\approx\frac{\phi^2}{2dh\tau}.
\end{equation}
Using that $\phi$ is large, we find local minima at $\phi=(2r+1)\pi$. Inserting this result into the R-branch of Eqs.~(\ref{eq:delta},\ref{eq:psi}) yields an expression for the $\delta$-value. For large $\eps$, we see from Eq.~\eqref{eq:delta} that $\delta$ is inversely proportional to $\phi$, while we see from Eq.~(17) that $\eps$ is proportional to $\phi^2$. Thus, $\delta \propto 1/\sqrt{\eps}$.

In summary, these results show that in the limit of large coupling strength there is a region of amplitude deaths at very small coupling delays. The onset of amplitude death occurs at a value of the coupling delay that scales as $1/k$, whereas the width of the amplitude death region (in terms of the coupling delay) scales as $1/\sqrt{k}$. We note that the reasoning presented in this section is not limited to the example topologies discussed above, but holds for all degree-homogeneous networks. While additional topological eigenvalues with $|c|=d$ could exist in directed or bipartite networks, the corresponding tongues would narrow the channels of amplitude death by a factor, but would show the same scaling behavior.

\section{Amplitude death in heterogeneous network}
As the final step of our analysis we investigate amplitude death in degree heterogeneous networks (DHENs).
In particular we focus again on the limit of large coupling strength and small coupling delays that 
corresponds to the bottom channel computed analytically in DHONs. For DHENs comparable analytical calculations are not easily possible. 
We therefore explore the stability of these systems numerically. Figure Fig.~\ref{fig:Scaling} shows a scatter plot in which each point correpsonds to specific values of 
$\eps$ and $\delta$, sampled uniformly, and a random Erd\H{o}s-Reny\'{i} network with $N=10$ nodes and $K=15$ links. For each of these randomly generated sample networks we used the numerical method described in \cite{Luzyanina1996,Hoefener2011} to compute the stability. In the figure unstable samples where marked by a red dot, whereas stable regions remain white.

From the scatter plot we see that for sufficiently large $\eps$, no unstable networks can be found in a large region at high coupling strength and small coupling delays. This region corresponds remarkably well to the analytical results that have been obtained for DHONs.
In particular there is a sharp loss of stability at he bottom bifurcation line. For sufficiently high coupling strength almost all unstable samples fall into the tongues of instability generated by the eigenvalue $c=d$ in DHONs, whereas few fall into the tongues with $c=-d$. These samples can probably be explained by the random creation of bipartite networks, and would hence be virtually absent in larger networks.   
 
\begin{figure}
\centering
\includegraphics{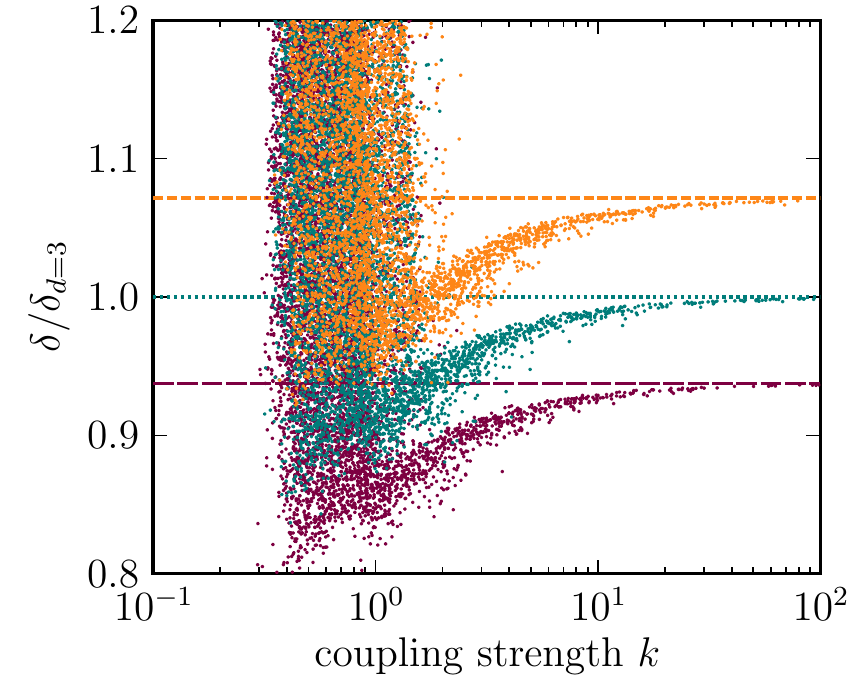}
\caption{Onset of amplitude death. Scatter plot of parameter values for which one randomly drawn network with $N=10$ nodes and $K$ links is stable and another one is unstable. Thus the dots mark the border between stable and unstable regions. The borders of network ensembles with $K=14$, $K=15$ and $K=16$ approach distinct lines, which are given by Eq.~\eqref{eq:delta_approx} with $d=2K/N$ (horizontal lines). The $\delta$-values are normalized by the line for $d=3$. Parameters are $\gp=-1,\lp=0,\tau=5$.}\label{fig:AvgDeg}
\end{figure}

As a further test we study the onset of amplitude death as the bottom (0,0) bifurcation line in more detail. To capture the onset of amplitude death we are interested in those systems values where for given values of $\delta$ and $\eps$ the stationary solution is stable
in one network topology and unstable in another topology. We thus repeat the sampling procedure above, but draw two topologies for every 
$\delta, \eps$ value pair. We discard all samples except those where the stationary state is stable in one topology and unstable in the other. Since we already know that the stability transition at the bottom bifurcation line is narrow for large $\eps$ we sample this region selectively by drawing $\delta$ from a bounded uniform distribution centered on $\delta_{d=3}(\eps)$, the bifurcation point for DHONs with degree $d=3$, which is given by Eq.~\eqref{eq:delta_approx}. This results in a uniform sampling of the parameter plane shown in Fig.~\ref{fig:AvgDeg}. We observe that for large $\eps$ the transitions occur only at specific values of $\delta$. These values very closely approximate the analytical predictions of the bifurcation points in DHONs, where we replaced the homogeneous degree by the mean degree
$d=2K/N$ of the respective DHEN. For large coupling strength the bifurcation line that governs the onset of amplitude death at small coupling delays thus seems to depend only on the mean degree of the network, whereas all other topological properties, at least in the Erd\H{o}s-Reny\'{i} ensemble, can be neglected. 

\section{Conclusions}
In the present paper we investigated amplitude death in general networks for delay-coupled delay oscillators. 
Building on a previous result we were able to study the regions for amplitude death analytically for degree homogeneous networks. 
In particular we considered the limit of large coupling strength for which we showed that all relevant 
bifurcation lines can be traced back to certain eigenvalues of the systems adjacency matrix. 
 
We showed analytically that in degree homogeneous networks regions of amplitude death exist at large coupling strength $k$. 
Specifically, we investigated a region for which the onset of amplitude death occurs already at very small values of the coupling 
delay that scales as $1/k$. The width of this region scales as $1/\sqrt{k}$ and thus is significant, at least in a logarithmic sense. 
It is remarkable that this region of amplitude death comes arbitrarily close to, but never reaches the case of undelayed coupling. 

The results obtained here for delay oscillators contrast with the well-studied case of Stuart-Landau oscillators \cite{Reddy1998}, which do not show amplitude death for arbitrary large coupling strength and arbitrary small coupling delays.
We can speculate that the strong self-feedback in delay oscillators might be in destructive resonance with the terms arising from the coupling to neighboring nodes. Thus, in contrast to Stuart-Landau oscillators a direct force working against the oscillatory dynamics might not be necessary. 

Numerical investigations indicate that the analytical results for degree homogeneous networks can be extended to the case of degree heterogeneous networks.
For sufficiently large coupling strength we found that numerical results for degree heterogeneous networks were in perfect agreement with analytical expectations for degree homogeneous networks. 
While these results were based on network topologies from an ensemble of Erd\H{o}s-Reny\'{i} random graphs one can assume that they should hold generally, at least for networks with exponentially decaying degree distribution.   

Regarding degree heterogeneous networks we noted that the onset of amplitude death for strong coupling and small coupling delays seems to depend on the networks mean degree, but not on other properties. This is remarkable as it shows that the node dynamics become sensitive to a global and hence delocalized quantity, which hints at a diverging correlation length. This can be made plausible by considering that strong coupling creates a stiff system that can rapidly communicate over the relatively small diameter of the networks. Furthermore, we have seen in the degree homogeneous networks that the dominating instability in the limit of strong coupling are in-phase oscillations, which are an inherently global phenomenon.      

\bibliography{bibdelay_edit_abbr}

\end{document}